# SARS-CoV-2 Impact on Online Teaching Methodologies and the Ed-Tech Sector: Smile and Learn Platform Case Study

**Natalia Lara Nieto-Márquez[1,2], Alejandro Baldominos[3,4], Almudena González Petronila[2]**
[1]Faculty of Health and Education, Universidad Camilo José Cela, 28692 Madrid, Spain
[2]Educational Department, Smile and Learn Digital Creations, 28043 Madrid, Spain
[3]Computer Science Department, Universidad Carlos III de Madrid, 28911 Madrid, Spain
[4]Data Science Department, Smile and Learn Digital Creations, 28043 Madrid, Spain

## Abstract

In the past few years, the use of new technologies and digital, educational material has been increasing. Owing to the situation brought about in Spain by the SARS-CoV-2 (COVID-19) pandemic, schools have closed and families were asked to confine at home since March 2020. During this period there has been a need of resources to be made available to the families and teachers so that they would be able to continue their teaching practice. Consequently, this study analyzes the importance of online methodologies and usage tendency of an educational resource example: The Smile and Learn platform. Thereby, the study presents the different models implemented to support education and its impact in the use of the platform. The analyzed outcomes and their effect on education are discussed.

## 1. Introduction

For many years, there has been observed in the education sector an increase in material, platforms and online resources (Brown and Green, 2017). This may indicate a decision to invest in implementing digital resources in the classroom. Nevertheless, the process of implementing such resources has been challenged by limiting factors. Some of these limiting factors related to technology can be the lack of devices in schools, obsolete devices or poor connectivity. Regarding teachers, who in many cases they are motivated to implement this sort of resources, the lack of specific training and time to be able to use them in class are highlighted (Bekker et al., 2015; McKnight et al., 2016). In many instances, this leads to inadequate use of online methodologies or digital devices.

In view of the global pandemic, the confinement and the closings of schools, the educational community has been obligated to rely on these resources (Crawford et al., 2020; Huang et al., 2020; Zhou et al., 2020). This necessity has been encouraged for the continuity of instruction. In this way, many teachers have seen themselves compelled to adapt to the situation to be able to continue teaching. At the same time, this has signified an added engagement for the families. The application of online methodologies has also been limited by the shortage of electronic devices at home or the lack of digital competencies.

To be able to continue with their classes, many teachers have opted for online methodologies or e-learning. E-learning is characterized by being a distance teaching and learning methodology based on the use of technology devices and digital resources (Karmakar y Nath, 2014). In like manner, online teaching allows the continuity of the use of mixed methods that were already being used in the classroom. This way, this digital material can be used as complementary material to other methodologies that make use of digital resources like the Flipped Classroom method (e.g. Roehl et al., 2013) or game-based learning (e.g. Moreno-Ger et al., 2008).

Consequently, this article puts forward the analysis of the use of the Smile and Learn digital material in Spain during the COVID-19 confinement period. On March 9th, authorities announced that as of March 11th student attendance in Spain would cease. The confinement of the population was declared on March 14th. This study aims at answering the following questions based on user categorization: Has the number of the users of the platform risen? Has the amount of digital resources usage time increased? Which time slot are more popular regarding the use of such digital material? What kind of digital, educational activities are singled out during the use of the platform?

## 2. Plataform Smile and Learn

Smile and Learn is an intelligent, digital platform in the Ed-Tech sector. The platform includes activities that cover core curriculum subjects from preschool to elementary school (3-12-year-olds). This way, the different subjects are merged together facilitating the use of a single resource which can adapt itself to all areas of study. Likewise, the platform helps students to work with other aspects like competences and cognitive skills such as content language integrated learning (CLIL).

Currently (May 2020), the platform has more than 5,000 activities in five different languages (Spanish, English, Portuguese, French and Italian), ergo these activities are divided into:

- Quizzes: activities with the objective to evaluate the theoretical knowledge gained while working with applications of a specific study area/core subject. This way, in this kind of activities game mechanics like information, questions, answers and feedback are interconnected with learning mechanics like identifying, discovering, tasks, reflection or repetition. The pedagogical objective in combining these mechanics lies in understanding, remembering, analyzing and evaluating (Arnab et al., 2014; Lim et al., 2015). Similarly, according to Baptista et al. (2015) using quizzes in learning promotes the development of strategic thinking skills like decision making (decision quality) and technical learning.
- Games: more practice-based activities to work on curriculum concepts or cognitive skills. In these activities, game mechanics like behavior, information, steps, answers, progress, points or feedback are interconnected with learning mechanics like guides, identifying, exploring, experimenting, tasks, repetition, reflection or analysis. Thereby, the pedagogical objectives are: to understand, remember, analyze, apply and evaluate (Arnab et al., 2014; Lim et al., 2015). The games integrated in the platform work on strategic skills like Technical Learning, Decision quality and problem solving, Intellectual Acumen. These games alsos work on operational skills like the amount of time spent in decision making. (Baptista et al., 2015).
- Stories: activities to foment an interest in reading through interactive resources and reading comprehension (Rajabi y Khodabakhshzadeh, 2015; Wright et al., 2013).
- Theory: activities to learn basic concepts through interaction with the device
- Videos: through visual channels and with the help of theories like the Cognitive Theory of Multimedia Channels (Mayer, 2005), core curricular theory concepts are visually displayed for learners. By this means, what is intended, is to achieve a meaningful learning experience using different information processing channels (visual and auditory) (Zhang et al., 2006).

These types of activities have been arranged in eight different worlds according to different learning areas:

- Literacy: In this world students work on core curricular literacy content together with grammar-based concepts. There are other applications as well focusing on verbal-linguistic intelligence, stories with values, Escape Rooms or reading comprehension skills.

- Logic: it features core curricular content related to the subjects of math and robotics. There are also activities for learners to work on their calculation skills, thinking process speed, memory and reasoning.
- Science: this world features several categories aimed for learners to work on scientific thinking and science concepts.
- Emotions: this world features plenty of applications that help learners work on emotional intelligence. In like manner, there are also activities that discuss the responsible use of electronic devices aiming at raising awareness regarding this issue.
- Art: the categories in the Art World include activities to work on art and music core curricular concepts. At the same time, there are also activities for learners to practice graphomotor skills, planning and organization skills, and creativity working with art and drawing concepts.
- Spatial: this world features different activities arranged in categories for learners to practice visual-spatial skills, coordination skills, visual-motor skills and thinking process speed.
- Multi-Player: in this world there are activities that allow learners to play together using one device. This way teamwork and social skills are boosted by means of technology.

All activities provide learners with progress feedback with relation to the mistakes they made, the answers they got right and the time spent on the application. Some activities give a more personalized feedback regarding going up levels, reflection questions to evaluate learning or moving up to the next level (C.f. Lara Nieto-Márquez, Baldominos y Pérez-Nieto, 2020).

- The eighth world is customizable to each profile: this last world is related to the gamification system which is integrated in the platform. This world features two categories: Fantasy Village and Urban Village. In both words, users can work on resource management skills using the tokens they earned while playing (smilies) or the ones they earn every time they go up a level (gems).

The platform features an open source gamification system. Each time a child plays, he/she gains a number of tokens depending on the difficulty or time needed to complete the activity. Concluding the challenge will help the learner go up a level, based on the number of activities attempted. This way, learners will be able to earn "gems" when they go up a level. Gems and Smilies will be used to build cities in both Villages, in the eighth world. They can also use them to customize the avatar and their house.

The platform also features a recommender system based on artificial intelligence. The platform's lower tab serves as a recommending tool which suggests activities to learners based on their progress. They can try these activities in order to continue improving academically. The recommender system also suggests activities based on what learners may like most, in regard to what he/she played before (C.f. Baldominos y Quintana, 2019).

Smile and Learn has also collaborated in different proposal plans focused on all the population:

On the one hand, there has been an agreement with the Spanish Ministry of Education and Training, and Radio Televisión Española (Spanish public service television) according to which specific Smile and Learn didactic content (videos) will be broadcasted from Monday to Friday during the morning time slots. The morning show "Aprendemos En Casa" (Learning At Home) is an inclusive, educative program which aims at reaching every household, including those that lack interactive electronic devices.

Smile and Learn has also presented families with an offer exclusive for them, which enables free access to all platform content for the first month.

## 3. Results

Next, we display the increase in the number of users and the time spent using the platform according to the classification made by schools or users in general.

### 3.1. COVID-19 Impact on the usage of Smile and Learn

Concerning education centers and schools, there has been an increase in the need of digital material. Figure 1 shows the rising number of users which opted for the school promotional offer as of March 11, when the school closing announcement was made official.

Upon analyzing the nature of the subscriptions of general users (number and time of use) which cover families too, we have observed an increase in the number of petitions in the beginning of the confinement period.

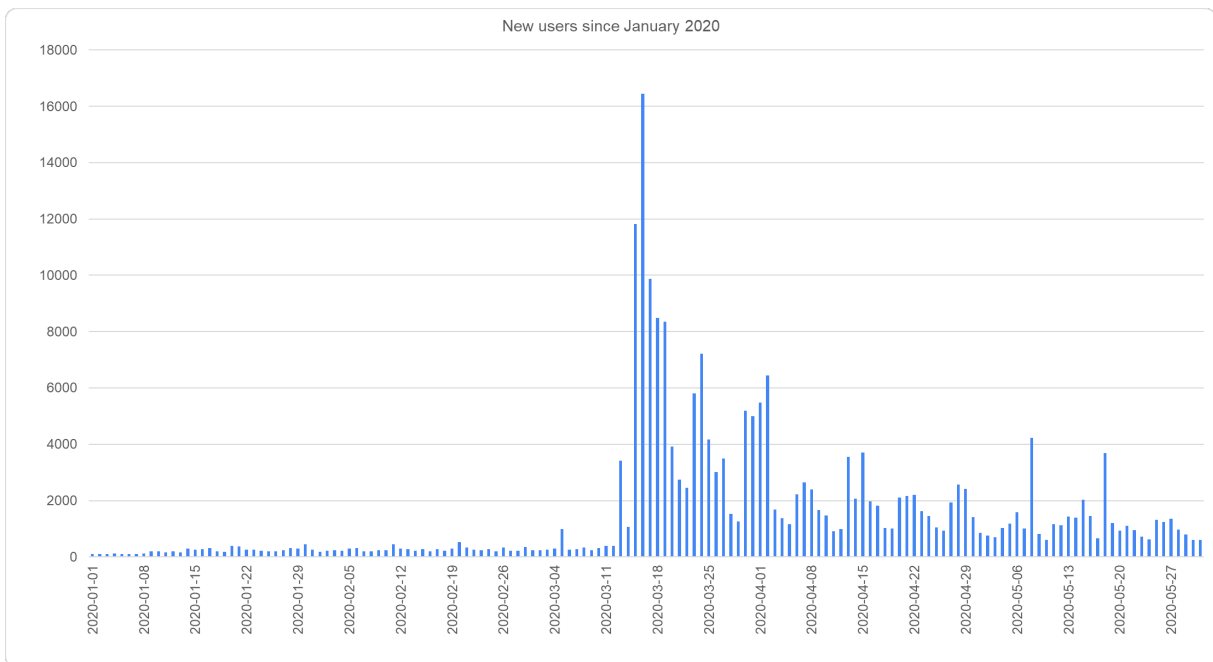

**Figure 1.** Users subscribed since January 2020 (Y axis is relative to 100%, which is the value for Jan 01).

### 3.2. Schools usage.

Regarding user accounts created for schools with the COVID-19 promotional offer, we highlight the peak observed one week after the confinement had begun. The number of school accounts starts dropping during the weeks following the confinement official announcement.

To be able to analyze the use of material by the schools, we have calculated the time schools employed using this material since January 2020 and the time schools that subscribed owing to the confinement situation, have started to employ. Consequently, Figure 2 reflects a daily use based on the sessions planned by the school, while Figure 3 reflects a more consistent use.

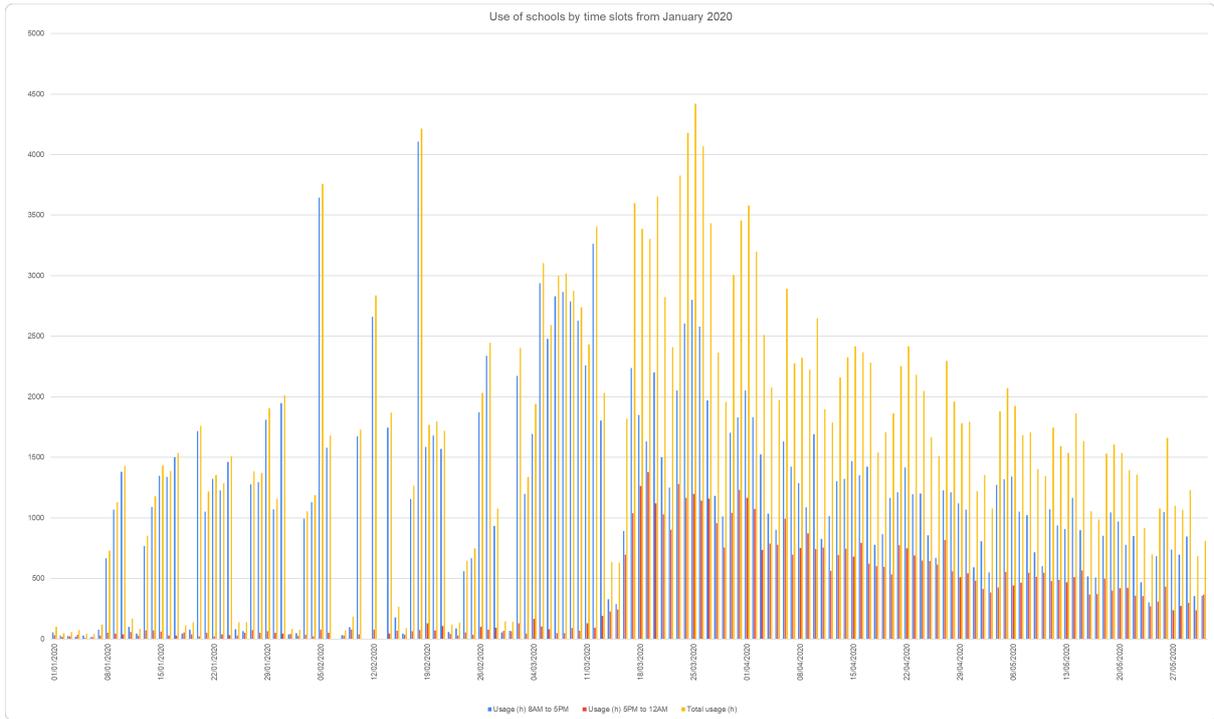

**Figure 2.** Smile and Learn Platform Use by Subscribed Schools - Time Slots Defined as of January 2020 (Y axis is relative to 100%, which is the value for Jan 01).

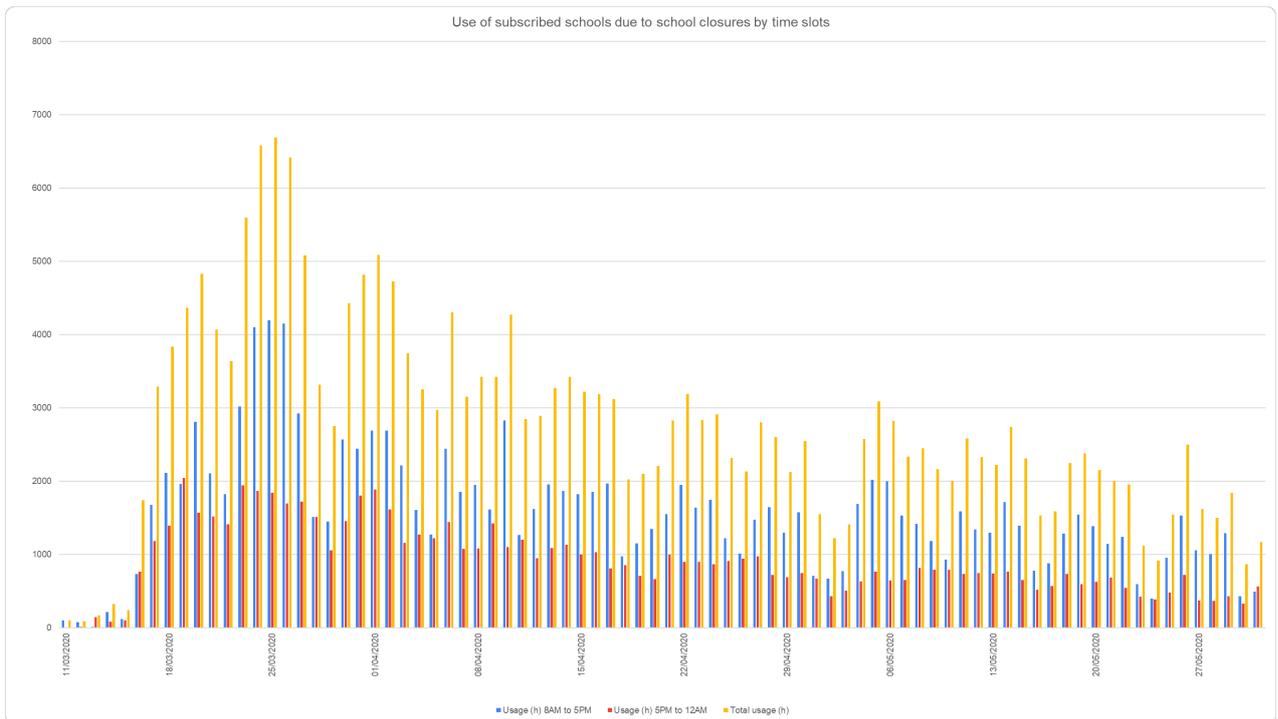

**Figure 3.** Smile and Learn Platform Use by schools subscribed owing to schools closings as of March 11, arranged in time slots (Y axis is relative to 100%, which is the value for March 11).

A significantly higher use of the platform has been observed between March 17 and March 18. This fact coincides with an increase in the number of subscription petitions received by schools. On another note, Figure 2 shows that the platform has been barely used outside school hours. Both Figure 2 and

Figure 3 reflect the fact that once the confinement period begun, there has been an increase in the use of the platform outside school hours. However, the use of the platform during school hours (from 8am to 5pm) is still greater.

Notwithstanding, regarding schools there has been observed a residual tendency of use during the weekends before confinement started. However, during confinement there has been an increase during the weekends, in the use of the platform by users designated as "schools". Parents work during the week and there are times that they cannot attend to the needs of their children. That's why certain tasks could be scheduled for the weekend. In like manner, the increased use of the platform can be explained by the fact that families have been more involved in the learning process and support of their children. As a result, depending on the situation of the parents, study time slots for the entire week may be organized accordingly.

With regard to the type of activities used in the platform, we highlight the use of those classified as educational games. Figure 4 depicts activity use in percentages: games, theory, questionnaires, videos or stories.

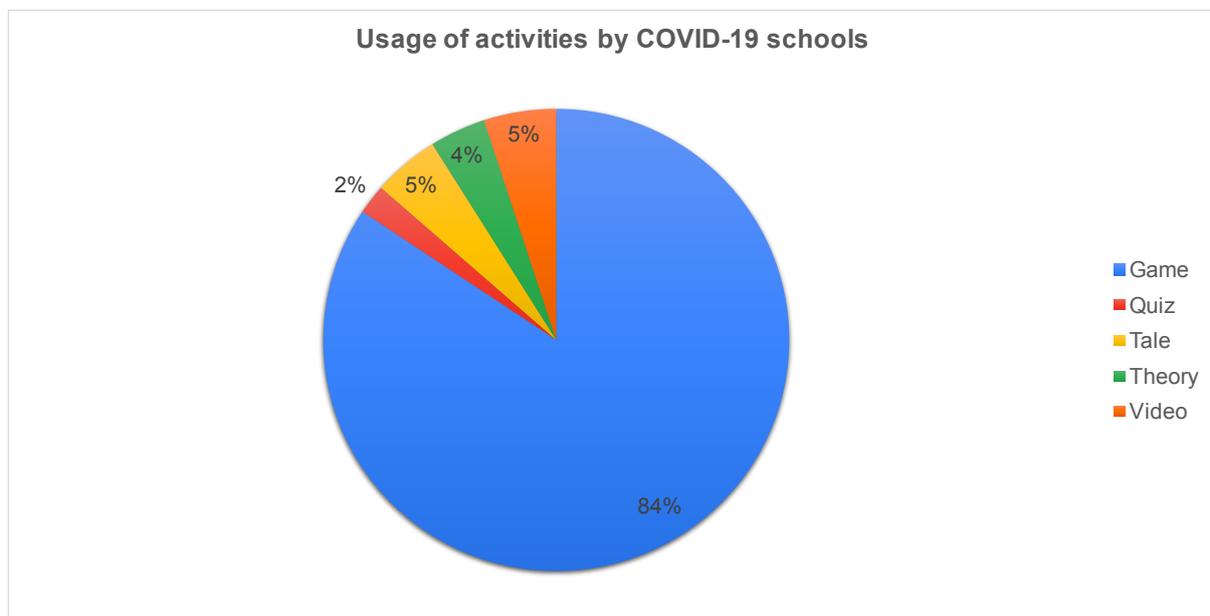

**Figure 4.** Percentage of time of use of the activities by educational centers during the months of March, April and May.

Following the use of games, videos are the second choice/option for students to practice core curricular content.

### 3.3. General usage

In contrast to education centers, the use of the platform by general users is homogeneous, as it is evenly distributed along all time slots, aside the school hours limitation (Figure 5). Once the confinement period began, the number of users has increased same as the number of schools. This rise in the use of the platform mirrors the increase in use of technology of the families subscribed.

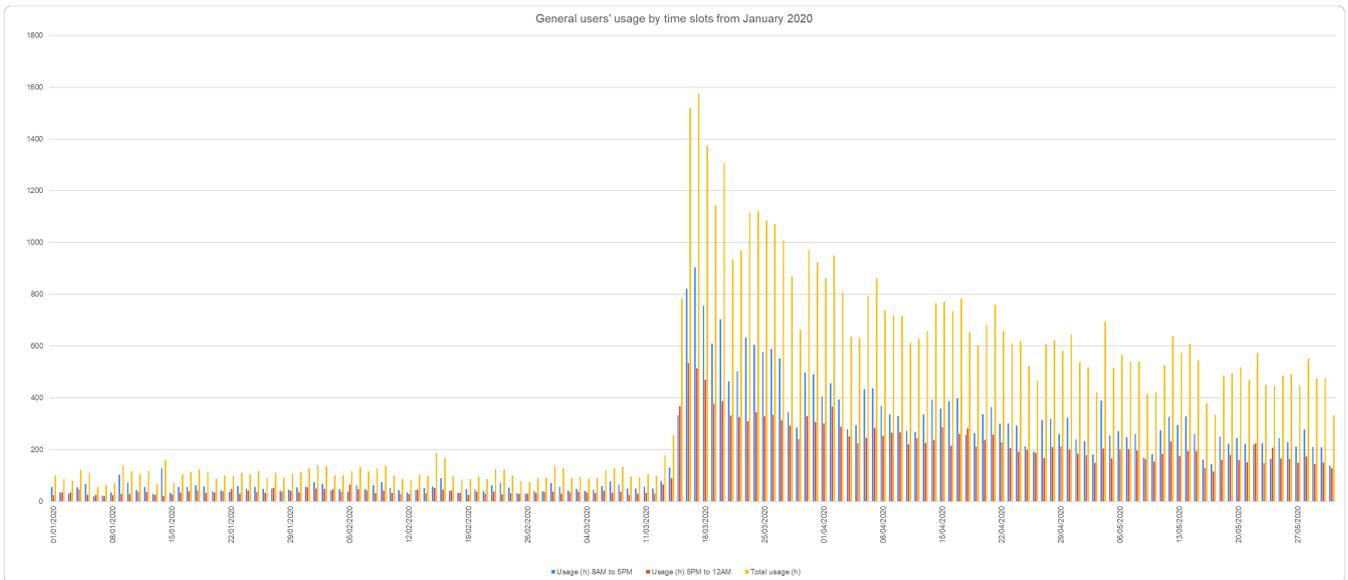

**Figure 5.** Use of the Smile and learn platform by general users in different time slots based on hours (Y axis is relative to 100%, which is the value for Jan 01).

In relation to the use of activities, games are still marked as the number one choice of the users, followed by videos (Figure 6).

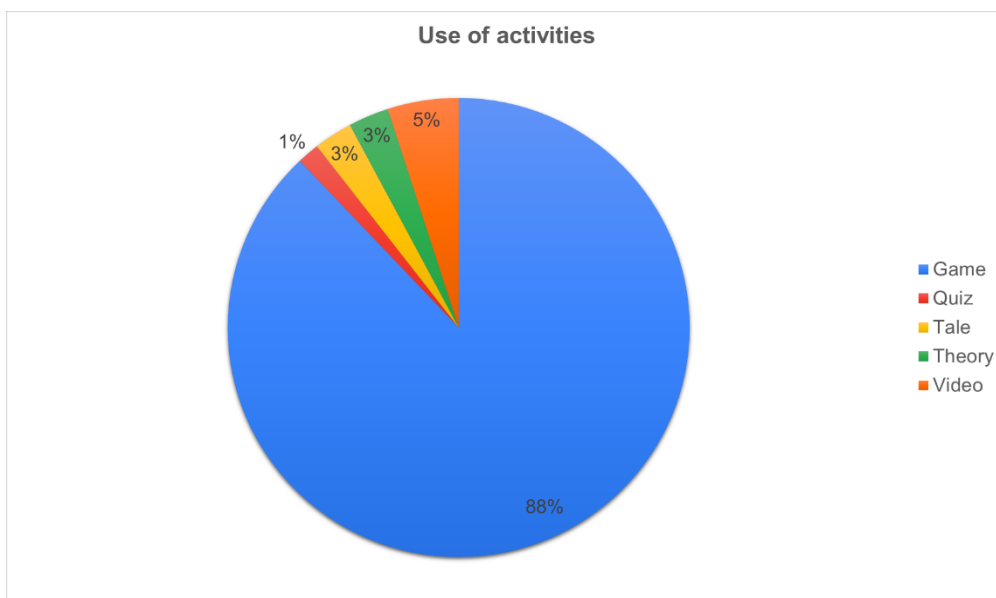

**Figure 6.** Percentage of the time of use of activities by general users in the months of March, April and May.

### 4. Conclusions and Outline of Action Plan

The use of activities and the number of Smile and Learn users has increased owing to school closings and the confinement of the families. This fact has brought about a greater need/demand of digital resources on behalf of teachers as to continue their classes with their students. In similar manner, there has been observed an increase in the general use of the platform, including weekends. This way, it would be interesting to ask ourselves whether or not there has been a change in the schedules of the students, and how such change has affected family life.

This study regards the utility of these digital resources as supplementary material to teacher practice and home based learning. The study also underlines the fact that the most popular learning activities according to the preferences of different users are games followed by videos.

In a similar way, we should take this investigation a step further and look into teacher training regarding those teachers who actually did use the Smile and learn resources, and the methodology they followed. We also stress that content customization makes more sense when the teacher is not present. Digital material enhances the possibility of tracing learning to boost areas students are challenged by. This task can promote ongoing evaluation.

Another course of future study to consider would involve the resources available to families apace with the use of this material by students. This way, we could direct our approach to the needs of the school community (families, teachers and students) in relation to the use of technology and digital educational material.

## 5. Acknowledgements